\documentclass{article}
\usepackage[a4paper, margin=3cm]{geometry}
\usepackage[style=apa,backend=biber]{biblatex}
\usepackage{moreverb,url,dirtytalk,graphicx}
\usepackage{authblk}
\usepackage[T1]{fontenc}      
\usepackage[utf8]{inputenc}   
\usepackage{lmodern}          

\begin{document}


\title{The Urban Model Platform: A Public Backbone for Modeling and Simulation in Urban Digital Twins}

\author[1]{Rico H Herzog}
\author[1]{Till Degkwitz}
\author[2,3]{Trivik Verma}

\affil[1]{City Science Lab, HafenCity University Hamburg, Germany} \affil[2]{Bristol Digital Futures Institute, University of Bristol} \affil[3]{School for Policy Studies, University of Bristol}



\maketitle

\begin{abstract}
Urban digital twins are increasingly perceived as a way to pool the growing digital resources of cities for the purpose of a more sustainable and integrated urban planning. Models and simulations are central to this undertaking: They enable "what if?" scenarios, create insights and describe relationships between the vast data that is being collected. However, the process of integrating and subsequently using models in urban digital twins is an inherently complex undertaking. It raises questions about how to represent urban complexity, how to deal with uncertain assumptions and modeling paradigms, and how to capture underlying power relations. Existent approaches in the domain largely focus on monolithic and centralized solutions in the tradition of neoliberal city-making, oftentimes prohibiting pluralistic and open interoperable models. Using a participatory design for participatory systems approach together with the City of Hamburg, Germany, we find that an open Urban Model Platform can function both as a public technological backbone for modeling and simulation in urban digital twins and as a socio-technical framework for a collaborative and pluralistic representation of urban processes. Such a platform builds on open standards, allows for a decentralized integration of models, enables communication between models and supports a multi-model approach to representing urban systems.

\end{abstract}


\section{Introduction}
Starting from the mere conception and perception of reality and ending at the details of software architecture, deployment and its embedding in social structures, the development of Urban Digital Twins (UDT) is an inherently interdisciplinary endeavor. Serving as a cross-domain knowledge broker for the purpose of integrated and sustainable urban planning, main areas of application include representation, analysis, communication and control \parencite{al-sehrawy_digital_2021}. Models and simulations are core components to enable what-if scenarios, utilize ever-increasing amounts of data and generate new insights \parencite{dalibor_cross-domain_2022}.

Although much seminal research has been conducted to represent various kinds of urban processes and phenomena in computational models, these are usually thematically focused and do not enable interactions with other models. Examples run the gamut: sophisticated models in traffic and mobility \parencite{smolak2020population}, land use interactions \parencite{forrester1970urban}, physical phenomena such as urban heat islands \parencite{nakata2018tool}, flooding and air quality \parencite{badach2023spatial}, or social processes such as demographic development and inequalities \parencite{nicoletti2023disadvantaged}. In existing UDT solutions and in the tradition of neoliberal smart city making \parencite{cardullo_smart_2019}, such simulation models are often engulfed in proprietary application wrappers that limit the use of interoperable (counter-)models both by their software design and governance choices \parencite{raes_duet_2021, dembski2019digital, schrotter_digital_2020}. There is a lack of publicly owned digital infrastructure to support the transparent and open usage of algorithms for planning and decision-making purposes.  

The conceptualization of cities as highly complex adaptive systems and the need for integrative planning processes in urban development call for pluralistic multi-modeling practices \parencite{batty_multiple_2021}. As structural and parametric uncertainty affect the results of every simulation model, multiple models constructed by multiple stakeholders are one possible way to mitigate each other's blind spots and shortcomings by analyzing differences and concordances \parencite{walker_defining_2003, page_model_2018}. To enable multiple simulation models, the possibility for co- and counter-modeling \parencite{greenberger_models_1976}, as well as to prevent path dependencies and vendor lock-ins in software infrastructure, public focussed socio-technical approaches are needed. Keeping people at the center of technology design, we follow a modified participatory design of participatory systems approach as a methodology to address the challenges and issues with current ways of smart city making. This practice is conducted together with the City of Hamburg to both highlight the importance of publicly owned digital infrastructure and the use and maintenance of such infrastructure by local governments. As a result, we present an open Urban Model Platform as a public socio-technical backbone for modeling and simulation within UDTs. We describe both the software architecture built on open standards and existing technologies, as well as its embedding in emerging social and governance context. 

This paper is structured as follows: First, we lay down the theoretical basis of modeling and simulation in urban digital twins and describe the socio-technical embedding of current approaches. Next, we describe a participatory design for participatory systems approach. Following that, we present the results of a process that lasted more than three years of research through design and deliberate on the findings. Finally, we provide suggestions for future research.

\section{Background}
As a recent development in the context of \say{smart} cities, UDTs have become a sought-after tool for steering towards more resilient and sustainable cities \parencite{therias_city_2023}. However, there is still considerable ambiguity around the term digital twin, which generally describes some sort of digital system in a more or less direct relationship with a real-world system. Based on the degree of automated data flow between the \say{physical} and the \say{digital} counterparts, multiple scholars distinguish between digital models, digital shadows and digital twins \parencite{fuller_digital_2020}. While digital models require manual translation of data between physical and digital systems, digital shadows automatically receive data from the physical system through various means, such as sensor technology. Digital twins eventually close a feedback loop by automating data exchange between both parts \parencite{sepasgozar_differentiating_2021}. This allows the digital twin to be equipped with various algorithms for optimization, predictive maintenance, or other decision support use cases, and transfer optimized parameters to its real counterpart. Eventually, the physical and digital systems become deeply interwoven and almost indistinguishable \parencite{sepasgozar_differentiating_2021}. 

Models and simulations are an integral part of every digital twin. \textcite{dalibor_cross-domain_2022} find that in more than 75\% of the publications, the authors explicitly mention models as part of the digital twin. Such models can be of various categories: geometrical models, 3D models, simulation models, physical models, data-driven models, statistical models, etc. In the context of this paper, we view models as formalized systems of natural systems that are established through a modeling relation for the purpose of understanding and anticipation \parencite{rosen_anticipatory_2012}. If the modeling relation holds, the encoding of a natural system (N) into a formal system (F) allows to inquire within the latter and decode and interpret the insights back to the former. In case F describes state transitions based on an index variable (such as time), we define such a model as a simulation model \parencite{nance_time_1981} which allows for (virtual) experimentation and the creation of \say{what-if} scenarios, thereby forming the basis of digital twins. 

Although precise simulation models are achievable in certain domains of the city, scholars point out that a single multidimensional city-scale digital twin with reciprocal automated data exchange is impossible due to the underlying complexity of urban systems \parencite{nochta_socio-technical_2021, caldarelli_role_2023} and the serendipitous nature of life in complex societies. Thus, models and simulations in UDTs require different approaches to formalize their corresponding natural systems compared to digital twins developed for manufacturing \parencite{caldarelli_role_2023}. The distinctive feature of cities is the self-organization inherent to many urban systems, which gives rise to complex adaptive properties such as non-linearity, emergence, path dependency or system nestedness \parencite{holling_understanding_2001, dam_agent-based_2013}. Based on \textcite{rosen_anticipatory_2012}'s insights and the dichotomy between natural and formal systems, the complexity of a natural system can be defined as the number of independent modeling relations that are established for its investigation. In this sense, the complexity of N becomes closely related to the one(s) who observe the system and form the modeling relation(s). It also implies that for any sufficiently complex natural system, there are multiple valid ways of formalizing it. It is not to be mistaken with mathematical measures of complexity, which are inevitably linked to formal systems. Following these considerations, a central question in the development of UDTs emerges around supporting a pluralistic formation and integration of multiple modeling relations \parencite{batty_digital_2018, batty_map_2019}. This especially holds true in the context of neoliberal city-making where tech corporations and consultancies push for an \say{objective} truth in data and models, which they develop and sell \parencite{cardullo_smart_2019, cugurullo_rise_2023} -- although such lofty visions are mythical in nature.

The integration of various simulation models within Urban Digital Twins (UDTs) has been a focal point for both scholars and practitioners, leading to a range of conceptual and technical approaches. A key consideration in these integrations is the central role of three-dimensional spatial data, as emphasized by \textcite{schrotter_digital_2020}. Their study of Zurich highlights the importance of linking diverse simulation models and applications individually to a 3D representation of the city, ensuring a spatially coherent framework. Similarly, \textcite{dembski2019digital} conceptualize the digital twin of Herrenberg, Germany, as a \say{container for models, data, and simulation}, serving as an open-source visualization platform that integrates space syntax, pollution, airflow, and mobility models. Beyond these spatially anchored frameworks, other scholars have proposed more distributed and modular approaches to simulation model integration. \textcite{raes_duet_2022} introduce the concept of a \say{cloud of models}, wherein dockerized models are orchestrated by a central message broker. Their proof of concept, applied in Athens, Pilsen, and Flanders, demonstrates how this \say{T-Cell architecture} facilitates dynamic interplay between traffic, air quality, and noise emission models, enabling complex what-if scenario analyses. In a similar vein, \textcite{lohman_building_2023} propose an \say{Inter Model Broker}, a system designed to enable communication between independent modeling frameworks. This broker relies on in-memory data storage, allowing multiple simulation modules to operate within a unified framework in real time. For strategic, long-term city planning, \textcite{nochta_socio-technical_2021} advocate for a \say{suite of specialized models} tailored to different scales of urban decision-making. Their framework, developed for Cambridge, integrates existing sectoral models while designing specific connectors to facilitate their coupling. This approach underscores the need for flexible, adaptable integration strategies that accommodate diverse urban planning objectives over extended time horizons. Together, these studies illustrate the diverse methodologies employed in integrating simulation models within UDTs. Whether through centralized spatial representations, cloud-based modular architectures, or specialized strategic planning suites, researchers continue to develop solutions that enhance the interoperability, scalability, and applicability of digital twins in urban environments \parencite{schrotter_digital_2020, dembski2019digital, raes_duet_2022, lohman_building_2023, nochta_socio-technical_2021, BELFADEL2023101989}.

While being useful for numerous use cases in urban planning, such approaches to UDTs are typically centered around a single entity that creates, hosts and manages a city's digital twin in a monolithic way. Its incorporated models and simulations, including their implicit and explicit assumptions, are prone to reflect the aspects that the entity owning (or selling) the twin deems necessary. By design, a pluralistic formalization of any urban system is hindered. Depending on its concrete implementation, this may come with the cost of underestimating the performative effects of simulation models \parencite{thompson_escape_2022, caldarelli_role_2023} and the risk of their prophetic instead of predictive usage \parencite[p. 198]{deutsch_beginning_2012}. With the increasing usage of AI in city-making, driven by proprietary interests and the promise of more efficiency, such models are given more and more agency, hence they essentially \say{determin[e] what is right or wrong, fair or unfair, good or bad} \parencite[p.6]{cugurullo_rise_2023}. 

Urban and complexity science scholars reject such top-down control room approaches and promote the inclusion of participatory approaches and co-evolutionary practices \parencite{caldarelli_role_2023}. Following \textcite{rosen_anticipatory_2012} definition of complexity, multiple models capturing different facets of the same natural system (or capturing facets of the same natural system differently) are increasingly described as a useful way forward to integrate the complex properties of cities into their digital twins \parencite{batty_multiple_2021, page_model_2018, Fishwick1994multimodel}. Co-modeling with stakeholders and counter-modeling algorithms from different perspectives allow for a pluralistic digital representation of urban processes. Building upon these ideas, a UDT can also be conceptualized as a modular system, combining multiple data sources, (simulation) models and applications in a use-case-specific digital representation \parencite{schubbe_urbane_2023, Gil_Redefiningurbandigitaltwins2024}. However, there are currently no public and open socio-technical systems in place which are designed to enable multi-model representations of the underlying complex urban systems on a city-scale. The interoperability of models remains \say{the most severe challenge} \parencite[p.1]{lei_challenges_2023}. Hence, a socio-technical practice that is structurally capable of incorporating multiple decentralized simulation models in an interoperable way is necessary for the further advancement of digital urban twins.

\section{Approach}
To address both the lack of a technological basis for multi-modeling in the public interest in the context of UDTs and its embedding in sociopolitical planning processes, we follow a participatory design of participatory systems (PDPS) approach \parencite{van_langen_participatory_2023}. Design research -- more specifically research through design -- \say{is an approach to scientific inquiry that takes advantage of the unique insights gained through design practice} \parencite[p.1]{godin_aspects_2014} and allows inquiry into the ought-to-be through the creation and testing of (socio-technical) artifacts \parencite{frayling_research_1993}. By combining design practice with participatory approaches and complex systems theory, PDPS specifically takes into account the social system around technological artifacts. This entails the co-design of sustainable collaborations and practices together with different stakeholders, the alignment of values and the co-production of knowledge. Given the drawbacks of current modeling and simulation practices within UDTs and the exploratory research objective, we see such an approach as suitable for the investigation of novel socio-technical systems which enable a collaborative and pluralistic representation of urban processes. PDPS consists of five interchangeable and iterable phases, namely \say{setting the scene}, \say{exploring the current situation}, \say{defining possible futures}, \say{transitioning} and \say{consolidating the transition} \parencite{van_langen_participatory_2023}. During the intertwined research and design process, different modes of documentation enable a systematic inquiry. In this paper, we draw from a total of 12 relevant meeting protocols (M1-12), six interviews (I1-6) with city officials, two workshop protocols, one technical and one governance report, as well as the resulting codebase as the main design artifact.

\section{Research process}

\begin{figure*}[h]
    \centering
    \includegraphics[width=1\textwidth]{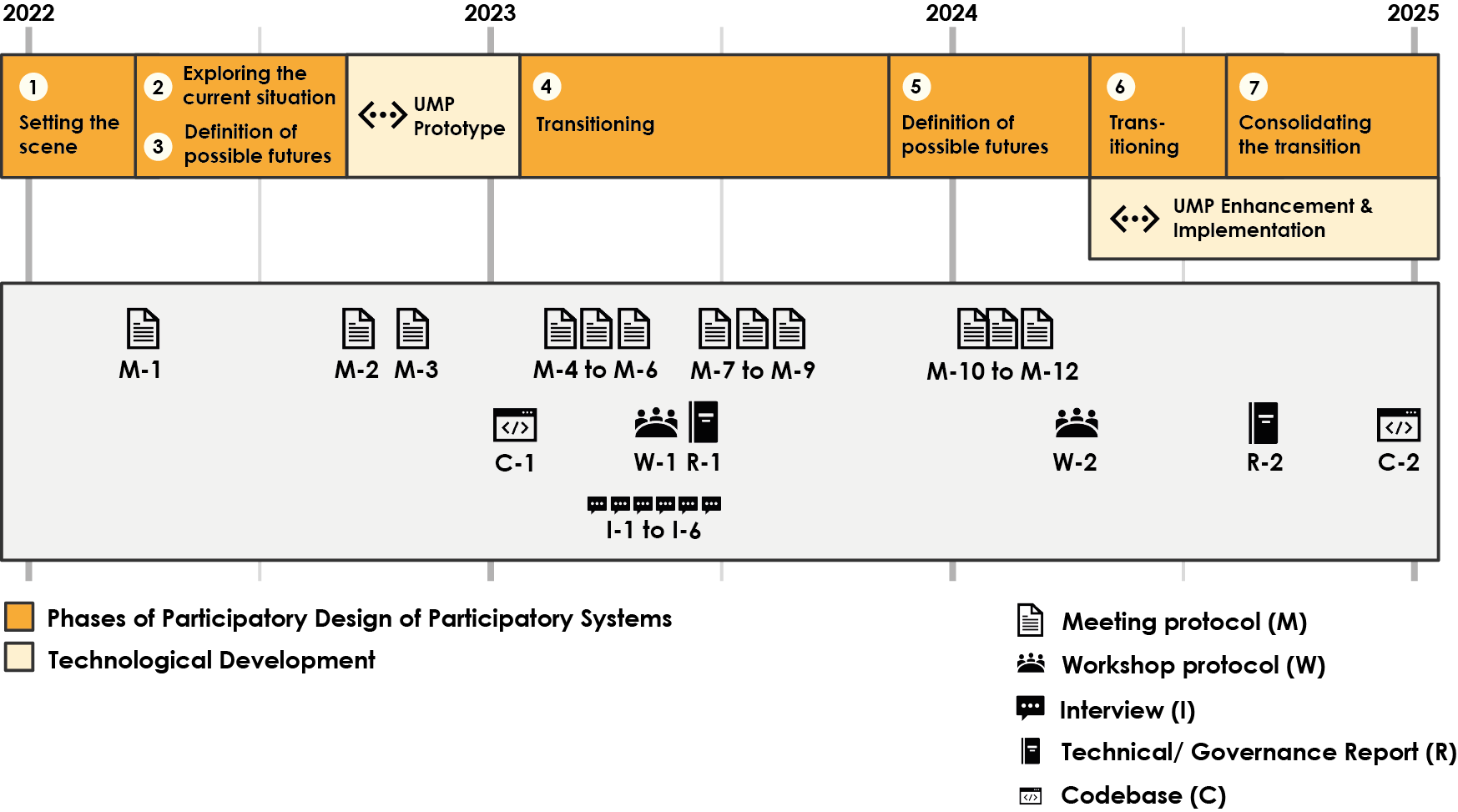}
    \caption{The research process spans three years and consists of seven PDPS Phases, two technological development phases and is documented by meeting protocols, workshop protocols, interviews, technical and governance reports, as well as the resulting codebase.}
    \label{fig-research-process}
\end{figure*}

As displayed in Figure \ref{fig-research-process}, various phases of our PDPS approach cover approximately three years from the beginning of 2022 until the beginning of 2025. Setting the scene took place in the context of a large-scale smart city project, where the City of Hamburg's Agency for Geoinformation and Surveying (LGV) and the City Science Lab of HafenCity University Hamburg continued a multi-year collaboration to create technological standards and processes for the city's digital twin project. Hamburg is Germany's second largest city and has established a comparatively advanced culture of data sharing, processing and usage in its decision-making processes \parencite{tegtmeyer_digital_2022}. This was largely achieved by leading efforts to standardize open urban platforms for data publishing and sharing \parencite{dinspec91357_openurbanplatforms2017}. The research partners explored the current situation in the form of surveys and during various exchanges with the LGV. The definition of possible futures started in February 2022 when LGV's officials shared their ideas of a city-wide model registry and open standards for transferring models from academia to the city's infrastructure. At this point in time, the discussion revolved around a cloud-based API to provide a single access point to a number of city-hosted simulations, which largely followed dominant UDT paradigms at the time (M-1). 

Broadening this concept with considerations of multi-modeling urban complexity, the research team prototyped an urban model platform at the end of 2022 (C-1). This artifact proved the technological feasibility of accessing and using multiple models created and hosted by different stakeholders, thereby creating a novel \textit{middleware} for models and simulations and applying the ideas behind an open urban data platform to models and algorithms. In a transitioning process, the City of Hamburg tested the artifact at the beginning of 2023 and founded an informal and monthly simulation task force between the LGV, the Ministry for Urban Development and Housing, the Hamburg Port Authority and research partners to advance the use of simulations across institutions. In July 2023, the city decided to commit to the technological system. In a collaborative refinement of next steps and necessary features, governance-related issues such as authorization, authentication, cost management, and data protection were discussed and largely implemented over the course of 2024. Consolidating the transition, since the beginning of 2025 a city-wide Urban Model Platform is operated by the City of Hamburg and the process of integrating sectoral models and algorithms has started. The updated digital strategy of the City of Hamburg now mentions \say{simulations, for example on mobility movements and developments, wind or shadows, are to be utilized and bundled in an Urban Model Platform} \parencite[p.37]{freie_und_hansestadt_hamburg_digitalstrategie_2025}.

\section{Results}
We present both the technical and social/governance design choices that were eventually implemented. Wherever relevant, drawing from the research material shown in Figure \ref{fig-research-process}, we describe the reasoning behind the design choices that were made.

\subsection{Governance design}
Integrated urban development supported by modeling and simulation in urban digital twins is an inherently collaborative and social endeavor which implies transitioning towards novel processes of urban planning. In such a transition, the design choices which shape the socio-technical system are based on a common understanding of current systemic problems and joint conceptions of a desirable future. 

City officials described the prevalent planning system as inefficient, fragmented, and understaffed. Together with an exhausted workforce \say{at the limit of their capacity} (I-4), neoliberal principles such as outsourcing scenario making to external private consultancies have created a system of long tenders and fragmented responsibilities. As one interviewee put it, \say{the city procures an expert report. [...] Then there's a consultant or an engineering firm. They return a pdf. In all levels of the administration, you can find structures which deal with handling and validating such kind of reports} (I-5). Given diffused competencies in scattered administrative departments, knowledge exchange is sparse, but needed: \say{I think it's really important that you have the opportunity to discuss things across domains and work together. But I think that most people don't do that} (I-3). One explanation for limited collaboration is the lack of tools which are fit for purpose. One interviewee in the city's planning department stated that \say{there are many people working in administration who do really dull work, just because there is no other way. They print aerial images, paint something onto them and scan them again. Obviously, this is work which takes up a lot of time. If we have tools that make things easier, we save resources which we can use for better things.} (I-2).

It is in this situation where the administration aspires to reclaim agency: \say{The more ownership an administration has [...], the more it is able to act with sovereignty and impact} (I-2). This implies shifting away from repeatedly procuring expert reports on highly specific scenarios, but rather building in-house competencies and digital public infrastructure. One city official suggested: \say{let's say we can put out a call for tenders for the creation of a noise simulation [..] and in the end this noise simulation model belongs to the city [...] and can continue to be used. So you no longer commission a model creation for subsequent procurements, but only a further development of the existing model [...]} (I-3). At the same time, city officials broached the issue of transparency and accountability. Within the administration, \say{information should be made transparent, perhaps by stating that this algorithm can be used for this, but not for that. In other words, one should explicitly name what it is or is not suitable for} (I-4). Externally, \say{there are [...] decisions where [the citizens] would like to know which variants were considered and what the parameters were for arriving at the result [...]} (I-4). Internally, the ability to share creates opportunities for model evaluation: \say{The aim is to facilitate preliminary checks and to give more users access to such tools [models] so that this can be spread across several shoulders} (I-5).

To address the lacking collaboration between departments and specifically enable a transfer of (simulation) models from academia to the public sector, the administration early on shared the idea of a central \say{model registry}, hosting models in their infrastructure and distributing access to city officials (M-1). The prevalent opinion was that \say{most city officials would like to use models and [...] compare scenarios, but not create them. They are happy if researchers do that} (I-3). At an early stage of 'defining the possible futures', the API Processes of the Open Geospatial Consortium (OGC) was named as a potential open standard with which models could be accessed and run (M-2). When presented with the prototype of a platform where multiple models can be connected from different stakeholders instead of providing all models on their own infrastructure, city officials acknowledged \say{valuable groundwork} (W-1) and later stated that \say{the model platform and subsequent scenario exploration can grow into something really big, which currently exists nowhere in this form} (I-3). At the same time, planning officials mentioned a more pragmatic approach in light of current decision-making processes: \say{If everything happens on a platform, that's nice to have. But for many things, [...] one can use established processes in the administration to coordinate matters and exchange. One does not have to do this on a platform. But of course it would be nice} (I-5). This was paired with skepticism that new technological approaches could not lead to a change of current decision making cultures: \say{I do not think that [..] a steering committee [...] would use a platform. I believe they will write their protocols, will use expert reports as PDF attachments, put them into emails, and file them in government folders.} (I-5). A high-level decision maker addressed the trade-offs involved in cultural shifts: \say{How big can the jUrban Model Platform be so that a bunch of people are taken along? Because if it's a huge jUrban Model Platform, a few people might watch enthusiastically, but nobody will jUrban Model Platform after it. [...] If it's just a very small step, I think it has added value, but what's really innovative about it?} (I-6).

In this context, the decision to integrate the prototype of the urban model platform into the city's digital infrastructure was taken. In a modular urban digital twin approach, the provider of the city's urban data platform decided to also provide a central hub for distributing various kinds of (simulation) models. In the following transition process, the research group hosted a workshop with city officials to discuss potential governance choices (W-2). Its aim was to identify missing technological governance areas. Workshop participants agreed that they were related to authentication and authorization of users, cost management, data protection and scalability. They emphasized the need for a governance system specifically for model sharing on an urban model platform and pointed out links to other governance systems, such as the EU's AI Act and GAIA-X as an evolving data space solution. As such frameworks were described to operate on a high level of abstraction, the need for concrete, pragmatic governance choices was pointed out. Most importantly, officials saw the allocation of costs in a federated platform of models as a crucial step to operate a model platform on a city scale (W-2). After researchers, city officials and a specialized start-up developed an expert report on technical governance choices (R-2), a second software development phase focused on implementing key governance mechanisms into the platform's code base. After the full technical implementation of the platform into the city's digital infrastructure was finalized in the beginning of 2025, unanswered social design questions remain. They are largely related to judicially sound contracts, procurements, models and payments, as well as the overall management of algorithms and capacity building within the administration. Contrary to a waterfall approach to finalizing the design of the software before rolling it out, city officials decided to move forward iteratively, starting with the integration of detailed and largely standardized wind and noise models for urban planning purposes. At the same time, the updated digital strategy of Hamburg names the Urban Model Platform explicitly for the utilization and bundling of models within the city's federated system of urban digital twins \parencite{freie_und_hansestadt_hamburg_digitalstrategie_2025}. 


\subsection{Technical design}
Four distinct elements make up the core of the technical design: (1) the Open Geospatial Consortium (OGC) API Processes as an open standard and interface to connect the other elements, (2) model servers and processing microservices of various stakeholders to compute various kinds of simulation models, (3) an Urban Model Platform as a novel contribution that provides an access point to multiple servers in a system-of-systems approach and (4) various front-end services connected to the Urban Model Platform. One key consideration underpinning the technical design was to thoroughly separate data and logic so that the same algorithms can be accessed and used by different applications, clients and users. 

\subsubsection{OGC API Processes}
Released in December 2021, the OGC API Processes is an open standard to provide an access point for computations on a web server via the Representational State Transfer (REST) protocol \parencite{ogcapiprocesses}. It largely builds on ideas of a web processing service (WPS) which was originally created in 2005 to provide \say{pre-programmed calculations and/or computation models that operate on spatially referenced data} via a web server \parencite{ogcwps}. Subject of the OGC API Processes is a standardized interface to such computations. Contrary to the XML-based WPS, the API Processes builds upon the JavaScript Object Notation (JSON) format and thus aligns well with the practices of modern full-stack web development.

The core of the OGC API Processes describes seven endpoints that ought to be implemented in every instance: An Open API-compliant landing page (1), a list of conformance classes that the server instance adheres to (2), and a list of all available processes on that server (3). Each of these processes is assigned a specific processID, which can be used to access a detailed process description (4). Using the same processID, clients can post input data to another endpoint where the process can be executed either directly before the server's response or queued for later execution (5). In the latter case of an asynchronous process execution, a job is created to obtain both a job status info (6) and the results (7). For synchronously executed processes, the results are directly returned to the client and no job is created. Within each of these endpoints, there are further requirements on how to specifically provide e.g. detailed metadata about the required inputs, outputs and job status \parencite{ogcapiprocesses}. 

In itself, the OGC API Processes open standard is agnostic of the programming language used for its implementation. Any provider of web-based computations is able to either choose from existing implementations in different programming languages or to develop their own API Processes-compliant endpoint within their service. In a heterogeneous environment such as public administration, this functionality is highly desirable, as it enables participants to maintain their local IT infrastructure while simultaneously contributing to a shared urban model infrastructure.

In addition to these technical considerations, the decision in favor of the OGC API Processes was also taken based on the surrounding socio-technical system (M-1 to M-4). OGC standards are used as a backbone for Hamburg's Urban Data Platform (UDP) \parencite{Hernández_IOTSummit-An_interoperable}, thereby enabling a diverse range of planning tools \parencite{degkwitz_cockpit_2020} and capability extensions such as the real-time sensor data infrastructure \parencite{fischer_urban_2021}. OGC standards do not only make up a large part of this fundamental infrastructure, but are key to enabling collaboration in urban planning processes and have established a general practice of using open APIs in the city's digital infrastructure \parencite{freie_und_hansestadt_hamburg_digitalstrategie_2025}.

\subsubsection{Model Servers and Processing Microservices}
Model servers host one or more simulation models, provide computational resources to execute them, and return scenario results. As the development of a UDT relies on the integration of cross-domain models, multiple model servers are necessary to digitally capture various aspects and dynamics of a city. Application Programming Interfaces (APIs) enable communication with other servers, applications, and Urban Model Platforms. In that way, a federation of model servers hosted and managed by different entities can provide multiple models to a range of applications within UDTs. 

A single model server can contain one or more processing microservices that calculate scenario results. Typically, each processing microservice runs calculations in its own containerized operating system, programming language, and installed dependencies. A model server of a given entity is hence able to provide access to multiple simulation models. 

Model servers and processing microservices are required in a modular UDT architecture for multiple reasons. Sufficient computational resources must be available to calculate scenarios for a broad range of UDT users. While sometimes personal computers can be used to execute simulation models for some use cases, e.g. in modern web browsers or with specific software, more sophisticated models require dedicated resources currently only available on large-scale server infrastructures. For instance, numerically solving differential equations for fluid dynamic and climate simulations might take days or weeks on high-performance clusters depending on their spatial resolution. AI models require large amounts of available random access memory (RAM) and graphics processing units (GPUs) for the prediction of features based on deep neural networks. Distributing various simulation models on different servers improves the efficient allocation of the required hardware. Additionally, companies offering proprietary simulation models may want to ensure the protection of their intellectual property by calculating scenarios on their own protected servers or in confidential computing environments with so-called \say{clean rooms} where a proprietary model and confidential data meet without the possibility of access for any party involved (R-2). Lastly, given the interdisciplinary character of the urban domain, there is a need for an efficient and scalable way to provide simulation models to an increasingly growing user base of UDTs. By decentralizing the providers of simulation models to a large number of urban model servers, it becomes cost-efficient and easy to expand the set of models available to UDT users and applications. At the same time, given the broad availability of cloud-based resources and access to the internet, there is little gatekeeping involved in creating and deploying new model servers.


On a technological basis, we extended the open-source project pygeoapi with a SocketIO endpoint to serve as a model server and connect various processing microservices. Developed as a Python-based Flask webserver, pygeoapi already integrates all required endpoints of the OGC API Processes, as well as several options for job management and result data storage. Processing microservices, once connected to the pygeoapi instance, register their specific processes with the server including metadata and wait for execution requests. To run a range of available open-source models, we implemented dockerized processing microservices in three different programming languages: Python, Node.js for ECMAScript languages such as JavaScript and Java. For specific modeling paradigms, we added support for NetLogo, Mesa Geo and InsightMaker models. For further technical information, we refer to the GitHub repositories for the model server (https://github.com/citysciencelab/urban-model-server) and the processing microservices (https://github.com/citysciencelab/processing-microservice-python).

\subsubsection{Urban Model Platform}

\begin{figure*}[h]
    \centering
    \includegraphics[width=0.5\textwidth]{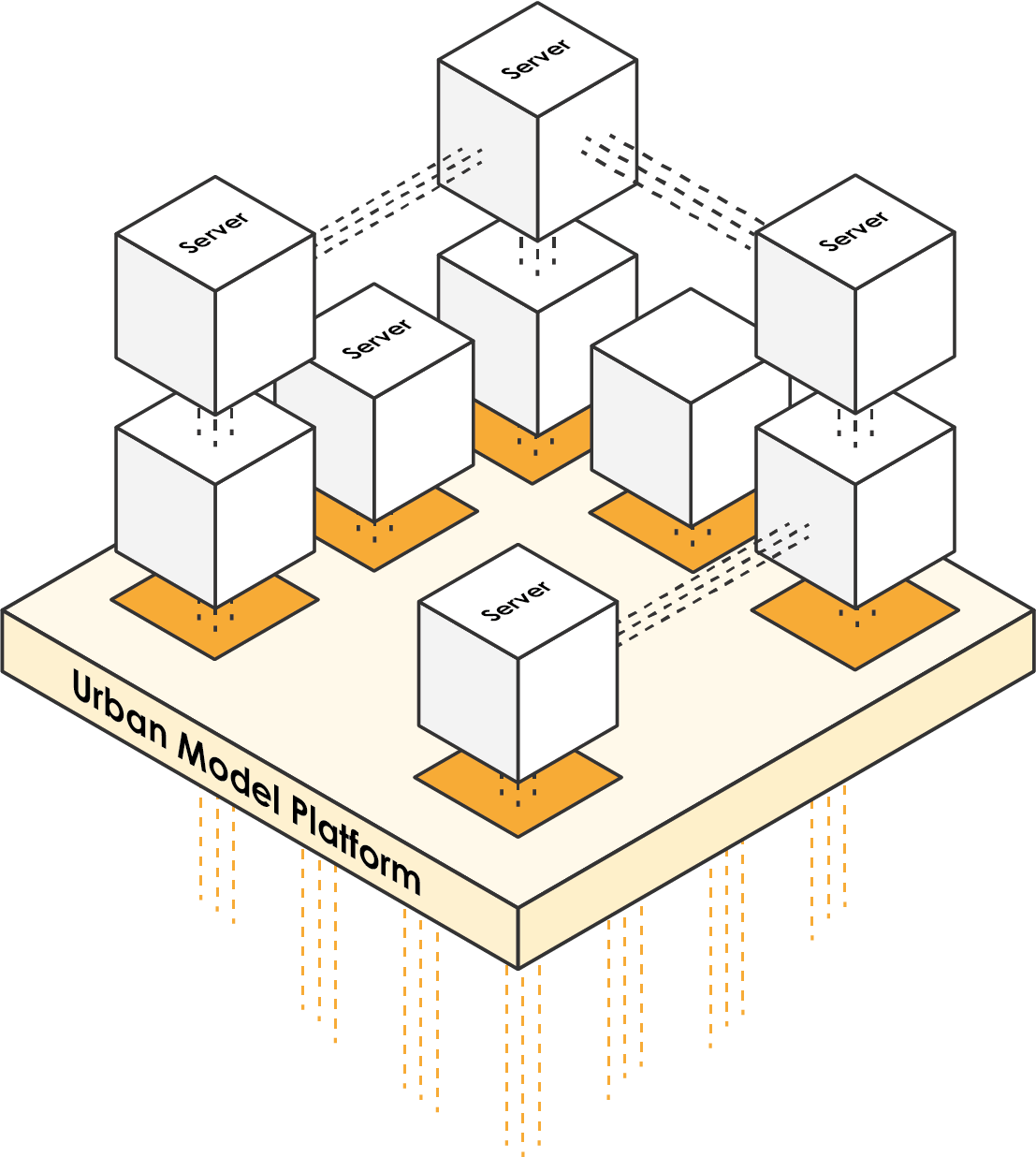}
    \caption{An Urban Model Platforms connect clients to one or more servers that execute simulation models}
    \label{fig:Urban Model Platform-schematic}
\end{figure*}

An Urban Model Platform as a novel technical artifact acts as an open urban platform for simulation models and algorithms provided by multiple model servers. Building on the standardized provision of process descriptions, execution, and result endpoints via the OGC API Processes, the Urban Model Platform can be configured to connect to multiple model servers and mirror their available processes in an OGC API Processes compliant way. Clients can retrieve information about available simulation models from a single point of access (i.e. the API of the Urban Model Platform) and send execution requests to the platform. The Urban Model Platform will then forward the simulation-specific input data to the correspondent model server. Once scenarios are computed on these servers, the results can be retrieved, stored, and provided by the platform (see Figure \ref{fig:Urban Model Platform-schematic}). This repository of scenarios can also be connected to an urban data platform and serve as a central access point and role management system to query and retrieve simulation results. Simulation models can rely on other(sub)models accessible via the Urban Model Platform, enabling communication between models and cross-domain model building.

A Urban Model Platform is by design not limited to a singular instance. In an environment of multiple available model servers, there can also be multiple coexistent model platforms managed by different entities (see Figure \ref{fig:multiple_Urban Model Platform}). This decentralized aspect limits the technological guardrailing of a single entity that creates, hosts, and manages a city's digital twin in a monolithic way. As some simulation models provide generalizable logic for varying input data, a single Urban Model Platform can also be connected to twin applications of multiple cities. One can envision the same simulation model of e.g. urban heat islands to apply to multiple cities if the correct input parameters are provided.

Due to varying skill levels, cost distribution and data protection, a truly cross-domain UDT with many stakeholders providing models and/or simulating scenarios requires flexible identity and access management. Both the permission to use computational resources for simulations and the access to the resulting scenario data need to be managed by the platform. 

Given these considerations, our implementation of the Urban Model Platform consists of different components to provide the capabilities outlined above. A Flask-based webserver compliant to the OGC API Processes provides the necessary routes and access points to the Urban Model Platform. In a central dynamic configuration file, the Urban Model Platform provider can register any number of model servers which are then mirrored in a standardized way via the platform's API. The provider of the Urban Model Platform can configure which models should be public, which should be in- and excluded, if the scenario data should be stored on model servers or mirrored on the platform and for how long the data should be stored. Additionally, we connect a PostgreSQL database and a Geoserver instance to log the various job requests and provide the results in multiple open standards such as Web Map Services or Web Feature Services. A Keycloak instance serves as an open-source identity and access management provider for different users and front-end clients. In addition to a fully dockerized architecture and the ability to host the platform on a scalable infrastructure, the City of Hamburg added Helm charts to run the Urban Model Platform in Kubernetes clusters. For further technical information, we refer to the GitHub repository (https://github.com/citysciencelab/urban-model-platform).

\subsubsection{Front-end Services}
Based on a thorough separation of logic and data, one design principle of the Urban Model Platform is to be agnostic of front-end applications. In this way, various applications can make use of the same underlying model. For instance, a single simulation model to calculate urban heat islands can be accessed from a web-based GIS system, a VR application or a mobile application. 

As one potential front-end service and to demonstrate the possibilities of a decentralized, multi-model approach, we developed a prototypical interface to the Urban Model Platform as an add-on for the open-source web GIS \say{Masterportal}. The Masterportal was originally developed by the City of Hamburg based on the JavaScript frameworks OpenLayers and Vue.js and is widely used across more than 45 German municipalities and regions, as well as state and federal institutions. Since there has been a growing stack of add-ons around the Masterportal's core \parencite{degkwitz_cockpit_2020, lieven_dipas_2017}, providing an add-on to communicate with the Urban Model Platform prototype technically enables a broad range of Masterportal users to access and execute multiple simulation models provided by the Urban Model Platform. This add-on named \say{Scenario Explorer} dynamically retrieves the required input parameters of the configured Urban Model Platform models and renders them in the user interface. Users can enter their parameter values and run simulations, as well as retrieve the results from finished scenario runs and display them in the web GIS interface.

\section{Discussion}

\begin{figure}[h]
    \centering
    \includegraphics[width=0.45\textwidth]{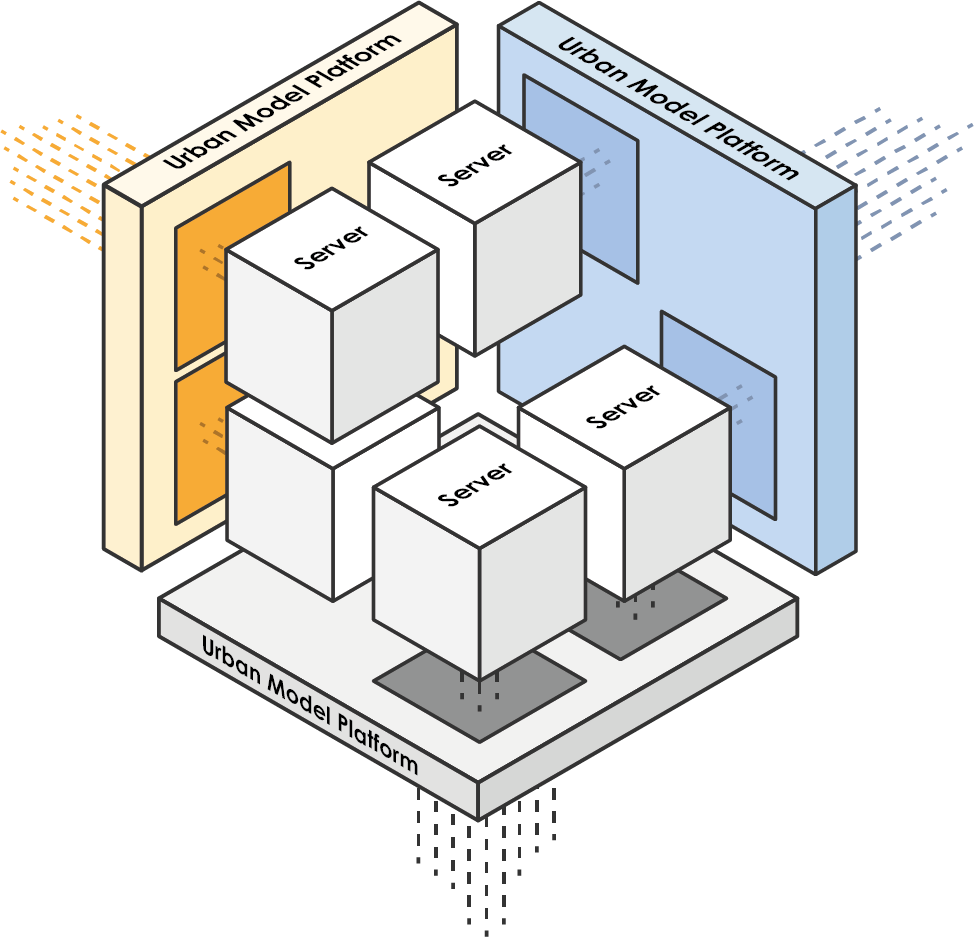}
    \caption{Conceptually, the proposed backbone can accompany multiple model platforms managed by different entities to prevent the control of a single entity over the city's urban digital twin}
    \label{fig:multiple_Urban Model Platform}
\end{figure}

By its design, the participatory system around the Urban Model Platform allows for the incorporation of multi-paradigm and multi-scale simulation models in a decentralized way and across multiple stakeholders forming a city's twin \parencite{nochta_socio-technical_2021}. Dedicated model servers and processing microservices execute simulation models across a broad range of formalisms, paradigms and programming languages. A system-of-systems Urban Model Platform connects the model servers with multiple front-end applications and with one another. Due to open and interoperable standards, a pluralistic representation of urban systems is not only technologically possible but also comparatively easy to implement and scale. As shown in Figure \ref{fig:multiple_Urban Model Platform}, one can envision multiple Urban Model Platforms that both prevent concentrating UDT infrastructure on a single entity or enable the connection of multiple twins. In this way, our approach differs from current solutions which either describe UDTs where models are depending on local and redundant data storage or encapsulate simulations in proprietary solutions. While the latter might be valid if the aim is to construct a single city-wide UDT \parencite{Deren2021}, we see a necessity for multiple twins deeply rooted in the reality of complex adaptive urban systems, planning practices with fragmented responsibilities and the resulting distribution of corresponding systems. It becomes possible to both share existent simulation models and algorithms between city departments, private planning bureaus and research institutions, as well as to counter-model dominant conceptions of urban systems \parencite{greenberger_models_1976}. 

Following the design ideas behind \textcite{dinspec91357_openurbanplatforms2017} and the socio-technical view of \say{gradually updating and expanding existing tools, processes and expertise} \parencite[p.283]{nochta_socio-technical_2021}, the proposed backbone builds upon a city's existent digital infrastructure and thereby facilitates the adoption by cities who are in the process of creating their digital twin. Existent tools can be upgraded to communicate with a Urban Model Platform which can improve their range of functions significantly. As the \say{Scenario Explorer} Masterportal add-on demonstrates, the web-GIS of over 45 municipalities and institutions is now technically able to execute and display the results of any number of simulation models available via a Urban Model Platform. Cities who choose to operate a Urban Model Platform for their digital twin can procure various simulation models separately and avoid vendor lock-ins that come with the procurement of a monolithic UDT solution. As mentioned by interviewees, such a socio-technical system contributes to a city's digital sovereignty, building  capacity inside the city administration and changing current processes towards more efficient ones. Additionally, simulation models can become more widespread and re-used within multiple applications. With the proposed technical backbone for a multitude of models and the ability to create multiple twins, collaborative practices of model creation are likely to evolve \parencite{herzog2023exploring}. Seen as a boundary object, the process of modeling and simulation can facilitate and intensify interactions between researchers, the modeled and decision makers \parencite{Boschetti2012Whatisamodel, cuppen_participatory_2021}. However, the extent to which this addresses more fundamental sustainability challenges in cities remains to be seen.

Given the parametric and structural uncertainties in simulation models of complex urban systems \parencite{walker_defining_2003, batty_multiple_2021}, a Urban Model Platform aids the process of integrating structurally different representations of the same urban system. UDT users are thus capable of comparing and contrasting results of multiple models to better evaluate structural uncertainty between models. This would pay homage to the complexity of city systems which can only be represented, simulated and ultimately understood through multiple lenses and paradigms \parencite{Batty2024}. For instance, established models, as well as novel AI-models could be deployed side by side. In the case of connecting models of various subsystems via a Urban Model Platform, it is necessary to pay close attention to compounding errors and/or their case-specific coupling. In the wake of increasingly agentic AI systems, we also see strong links to connecting the Urban Model Platform with generative AI models, providing promising starting point to a city's future AX design.

The resulting socio-technical system resembles multiple parallels to (urban) data infrastructures and urban data platforms (UDP), which continue to evolve into multi-actor ecosystems and data spaces. They form the core infrastructure for data-driven approaches that address contemporary urban challenges and smart city governance models \parencite{Barns2017SmartCA, bagheri2021value}. Building on this line of thought, a Urban Model Platform gives this premise a new maturity, as they are not independent of UDPs, but rather rely heavily on the integration of their data and building upon their existing networks and methodologies. Eventually, some version management system for actual and different predicted states of a city could help with integrating Urban Model Platforms and UDPs. Much like UDPs, the expected benefits of a Urban Model Platform arise from un-siloing, describing, organizing and structuring models and algorithms via a platform with a system-of-systems approach. Furthermore, while the introduction of urban data infrastructures has brought about new possibilities for integrated approaches to urban planning, it has also created an underlying necessity for increased trust among data-sharing entities, as well as in data-managing bodies, thereby fostering a data-sharing mentality within the public administration \parencite{tegtmeyer_digital_2022}. We believe that the introduction of Urban Model Platforms to the realm of urban modeling and simulation has similar effects in this field.

Multiple limitations and challenges come along with the socio-technical system design. From a methodological perspective, research through design is criticized to yield highly context-specific and hence non-reproducible results \parencite{godin_aspects_2014}. While acknowledging this shortcoming, scholars argue that research questions involving uncertainty and the inquiry of the ought-to-be typically operate in a conceptual space which needs to be explored. In such settings, the validity needs to be reconsidered. It can be established through a rigorous process design, via theoretical sensitivity or simply via the acceptance of the design artifact by stakeholders. Following this line of thought, we can report the latter, but also view a PDPS approach in designing the system as specifically powerful in addressing these questions together with practitioners. 

Concerned with more technological limitations, data and logic have to be thoroughly separated. This separation ensures that computations are always carried out on the latest available data and enables simulation models to generically function with multiple front-end applications. However, some models might depict local specifics with limited transferability. Such models require thoroughly calibrated parameters and are - by design - tuned to a specific context. Physical models, such as wind, noise or urban heat island models are, however, expected to bring a larger benefit to multiple city-wide or even cross-city UDT applications. As the ongoing process of forming a common ontology of urban objects evolves, we expect a modular and multi-model twin architecture to unfold its full potential. 

Another challenge is that it is not yet possible to run a single continuous simulation next to real-time with the inclusion of real-time data sources with the proposed back-end. Use cases involving \say{high-frequency city} \parencite{batty_digital_2018} are hence potentially difficult to include in the presented technological backbone as of now. This is largely due to missing standardized endpoints for a \say{continuous} execution mode next to (a)synchronous ones and missing options to real-time result queuing within the OGC API Processes. Since multi-modeling practices could also support \say{high-frequency city} use cases (i.e. running ensembles or multiple continuous simulations in parallel to predict traffic flows, pedestrian movement etc.), a future inclusion of a continuous execution mode is beneficial. Especially in the context of UDTs, the topic of including real-time data sources for modeling will likely become increasingly important, since they form the base for closely tying the digital and the real world and their interaction. 

Additionally, trust is required between various model providers to connect their models since a successful execution relies on the availability of the upstream model servers. With this, new forms of legal contracting between the different stakeholders involved are likely to develop. 


As with UDPs, the success of city-wide infrastructures for urban planning depends heavily on widespread adoption, the technical capabilities of the involved actors, and the willingness of stakeholders to share data — and/or models. Beyond the need to build trust and expertise within public administrations, another major challenge that Urban Model Platforms are likely to share with UDPs is the limited availability of data from non-public sources, including private companies and individual citizens. Incorporating these models is essential to accurately reflect and manage urban complexity. While open-source code and the possibility of interoperable, coexisting model servers offer a theoretical safeguard against these problems, they are unlikely to succeed without active facilitation and a strong commitment from public authorities. Without such support, Urban Model Platforms that represent the heterogeneity of the urban environment risk remaining purely theoretical.

\section{Conclusion}

Cities are highly complex adaptive systems. UDTs, aiming to make use of the underlying complexity by providing data and simulation models for more integrated and sustainable urban planning, should reflect this and incorporate multiple models of natural systems. In this paper, we introduce a Urban Model Platform as a socio-technological backbone for modeling and simulation in digital urban twins and outline a software architecture that builds on open standards, allows for a decentralized integration of models, enables communication between models and supports a multi-model approach to representing urban systems. In the surrounding social system, collaboration between stakeholders is fostered and urban processes can be represented in a pluralistic way while at the same time strengthening the sovereignty of cities. 

Multiple points for future research open up. First, the extent to which the potential for more sustainable development is being realized with Urban Model Platforms should be examined in further research. Second, research into semantic data models or a \say{version control} of different scenarios could ensure interoperable state management between models and scenarios.  Building on this, there is also a need to collaboratively design new interfaces that allow stakeholders to navigate, compare, and engage with multiple possible urban futures. Lastly, aside from the implementation of simulation models into UDTs, we see necessary future research in the participatory creation of simulation models in the urban domain. Such co- or counter-modeling practices would complement a multi-model technological backbone so that the diverse perspectives and values of citizens can practically find their way into an urban digital twin. 




\section*{Acknowledgements}
We thank Stefan Schuhart and Maja Richter of Hamburg's Agency for Geoinformation and Surveying for the co-development of the latest iteration of the model platform. Furthermore, we thank Pierre Gras and Michael Fischer for their invaluable knowledge regarding (real-time) data infrastructure, model registries and for providing starting points and advice for the present research. 

\printbibliography





\end{document}